\documentclass[a4paper,12pt]{article}
\usepackage{dirtree}

\usepackage{booktabs}
\usepackage{comment}
\usepackage{authblk}
\usepackage{tikz}
\usepackage{quantikz}
\usepackage[super,square]{natbib}
\usepackage{anysize}
\marginsize{2cm}{2cm}{1cm}{2cm}
\usepackage{fancyhdr}

\usepackage{soul}
\usepackage[dvipsnames]{xcolor}
\usepackage{hyperref}
\hypersetup{
    colorlinks=true,
    linkcolor=black,
    citecolor=CadetBlue,
    filecolor=CadetBlue,      
    urlcolor=CadetBlue,
}
\usepackage{lipsum}
\usepackage{multicol}
\usepackage{multirow}

\usepackage{parskip}
\usepackage{tabularx} 
\usepackage{amsmath}
\usepackage{graphicx}  
\usepackage{amssymb}
\usepackage{pifont}
\newcommand{\cmark}{\ding{51}}%
\newcommand{\xmark}{\ding{55}}%
\usepackage{forest}
\usepackage{xcolor}
\definecolor{foldercol}{RGB}{100, 100, 200}
\definecolor{filecol}{RGB}{50, 50, 50}

\usetikzlibrary{shapes.geometric, arrows.meta, positioning}
\tikzstyle{startstop} = [rectangle, rounded corners, minimum width=3cm, minimum height=1cm,text centered, draw=black, fill=blue!15]
\tikzstyle{process} = [rectangle, minimum width=3.5cm, minimum height=1cm, text centered, draw=black, fill=gray!10]
\tikzstyle{arrow} = [thick,->,>=stealth]

\renewenvironment{abstract}
 {\par\noindent\textbf{\abstractname}\ \ignorespaces \\}
 {\par\noindent\medskip}

\begin{document}
\pagestyle{fancy}
\thispagestyle{empty}
\fancyhead[R]{\textit{PCCC Galicia 2025}}
\fancyhead[L]{}
\renewcommand*{\thefootnote}{\fnsymbol{footnote}}

\title{\textbf{Benchmarking Distributed Quantum Computing Emulators}}

\author[1]{Guillermo Díaz-Camacho}
\author[1]{Iago F. Llovo}
\author[2]{F. Javier Cardama}
\author[1]{Irais Bautista}
\author[1]{Daniel Faílde}
\author[1]{Mariamo Mussa Juane}
\author[1,2]{Jorge Vázquez-Pérez}
\author[1]{Natalia Costas}
\author[2,3]{Tomás F. Pena}
\author[1]{Andrés Gómez}

\affil[1]{Galicia Supercomputing Center (CESGA)\\
Avda.\ de Vigo S/N, 15705 Santiago de Compostela, Spain}
\affil[2]{Centro Singular de Investigación en Tecnoloxías Intelixentes (CiTIUS)\\
Universidade de Santiago de Compostela}
\affil[3]{Departamento de Electrónica e Computación\\
Universidade de Santiago de Compostela}

\date{}  

\maketitle  

{\color{gray}\hrule}
\vspace{0.4cm}
\begin{abstract}
Scalable quantum computing requires architectural solutions beyond monolithic processors. Distributed quantum computing (DQC) addresses this challenge by interconnecting smaller quantum nodes through quantum communication protocols, enabling collaborative computation. While several experimental and theoretical proposals for DQC exist, emulator platforms are essential tools for exploring their feasibility under realistic conditions. In this work, we introduce a benchmarking framework to evaluate DQC emulators using a distributed implementation of the  inverse Quantum Fourier Transform ($\mathrm{QFT}^{\dagger}$) as a representative test case, which enables efficient phase recovery from pre-encoded Fourier states. The QFT is partitioned across nodes using teleportation-based protocols, and performance is analyzed in terms of execution time, memory usage, and fidelity with respect to a monolithic baseline.

As part of this work, we review a broad range of emulators, identifying their capabilities and limitations for programming distributed quantum algorithms. Many platforms either lacked support for teleportation protocols or required complex workarounds. Consequently, we select and benchmark four representative emulators: Qiskit Aer, SquidASM, Interlin-q, and SQUANCH. They differ significantly in their support for discrete-event simulation, quantum networking, noise modeling, and parallel execution. Our results highlight the trade-offs between architectural fidelity and simulation scalability, providing a foundation for future emulator development and the validation of distributed quantum protocols. This framework can be extended to support additional algorithms and emulators.

\end{abstract}
{\color{gray}\hrule}
\medskip
\begin{multicols}{2}
\tableofcontents
\end{multicols}
\newpage
{\color{gray}\hrule}
\section{Introduction}
{\color{gray}\hrule}

Quantum computing has emerged as a promising computational paradigm that uses the principles of quantum mechanics to tackle problems intractable for classical systems. However, the current landscape, commonly referred to as the Noisy Intermediate-Scale Quantum (NISQ) era, is dominated by devices with limited qubit counts and significant susceptibility to noise. One of the most critical challenges in this era is \textit{scalability}: building quantum computers of thousands or even millions of qubits that still maintain high coherence and fidelity. Inspired by classical Distributed Computing and High Performance Computing, this challenge has led to increased interest in distributed architectures that connect multiple smaller quantum processors to act as a unified, larger system. Distributed quantum computing (DQC) offers a possible solution by interconnecting multiple smaller quantum nodes via quantum communication links. This approach enables computational tasks to be split across nodes and coordinated using entanglement and classical control. Rather than scaling-in with a single and larger monolithic device, DQC enables modular architectures that could scale-out more efficiently across heterogeneous platforms and physical environments.

Distinguishing between DQC and the broader concept of the Quantum Internet (QI) becomes crucial. While both involve quantum networks, QI focuses on communication tasks, such as quantum key distribution (QKD), secure transmission, and entanglement distribution, whereas DQC is centered on computational collaboration across nodes. The goal of QI is to transmit quantum information; the goal of DQC is to process it cooperatively. QI wants to maximize the number of Bell pairs as an objective, DQC wants to minimize them as a resource. The boundary between these two fields is subtle but significant, and their technological needs differ in both protocol design and resource management.

DQC protocols are difficult to implement experimentally due to hardware constraints, node synchronization issues, and quantum channel fragility. As a result, \textit{software emulation} has become an indispensable tool for researchers. Emulators allow the prototyping of DQC protocols, exploration of system architectures, and simulation of control mechanisms and physical effects, long before scalable hardware is available. In this way, they accelerate algorithm development and offer insight into tradeoffs between noise, architecture, and performance. However, DQC emulators themselves vary widely in their capabilities and intended use. Some tools model detailed physical layers such as decoherence and photon loss; others focus on logical protocol correctness. Still others distribute the classical simulation effort across HPC resources to simulate larger quantum systems monolithically. This leads to an important distinction:

\begin{enumerate}
    \item \textbf{Distributed (or parallel) emulation of quantum computing:} using classical parallelism to simulate large quantum circuits on distributed memory machines (e.g., MPI-enabled backends).
    \item \textbf{Emulation of distributed quantum computing:} simulating protocols and behaviors of physically distributed quantum processors.
\end{enumerate}

This study focuses on the latter: benchmarking tools that \textit{emulate distributed quantum computing}. Specifically, we explore whether current emulators are capable of simulating non-trivial distributed algorithms, those that require teleportation, entanglement management, and inter-node gate execution. We categorize emulators into four quadrants based on their architecture (monolithic versus distributed) and their target functionality (monolithic QC versus distributed QC), as shown in Table~\ref{tab:emulator-classification}. To the best of our knowledge, no existing software framework provides both a faithful emulation of distributed quantum computing \emph{and} a distributed-memory classical simulation backend. Current tools cover one aspect or the other, but not both simultaneously.

\begin{table}[ht]
\centering
\begin{tabularx}{\textwidth}{|l|X|X|}
\hline
\textbf{} & \textbf{Monolithic Quantum Computation} & \textbf{Distributed Quantum Computation} \\
\hline
\textbf{Monolithic Emulator} & 
Qiskit Aer, PennyLane, myQLM, Qulacs, ProjectQ, IQS, QuEST [...] & 
NetSquid, SquidASM, SimulaQron, QuNetSim, Interlin-q, QuISP, SeQUeNCe,  DQC Executor [...]\\
\hline
\textbf{Distributed Emulator} & 
mpiQulacs, QuESTlink, IQS (with MPI), Qiskit Aer (multi-GPU) [...] & 
(??) \\
\hline
\end{tabularx}
\caption{Classification of quantum emulators by simulation architecture and computation model.}
\label{tab:emulator-classification}
\end{table}

To this end, we implement a distributed version of the Quantum Fourier Transform (QFT), a widely used algorithmic primitive in quantum computing, using gate teleportation to handle non-local interactions. We implement a distributed inverse QFT ($\mathrm{QFT}^{\dagger}$), rather than the standard forward QFT, because it allows us to prepare known input states in the Fourier basis and verify correctness by recovering the original phase upon measurement. This approach simplifies fidelity evaluation and avoids the need for complex state preparation across nodes. This benchmark serves to probe how well current emulators support distributed quantum circuits, measuring the computational and memory costs associated with emulating teleportation and coordination. The choice of the QFT comes from its importance for canonical algorithms such as the Quantum Phase Estimation or Shor's factorization algorithm, but also from its dense entanglement between qubits, which makes it hard to distribute. 

To our knowledge, this is the first benchmarking study focused specifically on the distributed emulation of quantum algorithms across quantum nodes. While previous works have benchmarked circuit simulators \cite{jamadagni2024benchmarking, Leonteva2025Apr}, they largely omit inter-node coordination, entanglement management, and protocol-level simulation. Our work highlights both the strengths and limitations of currently available tools in this area.

The remainder of this paper is structured as follows. Section~2 provides a comparative overview of existing distributed quantum emulators. Section~3 introduces the methodology of the benchmark task: a distributed inverse QFT based on gate teleportation. Section~4 presents experimental results on runtime, memory usage, and fidelity across several emulators and configurations. Section~5 discusses limitations, future directions, and opportunities for emulator development\footnote{The full benchmarking suite, including scripts, data, and plotting utilities, is available at \texttt{https://github.com/gdiazcamacho/DQC-Emulator-Benchmarking}.}.
\newpage
\section{Overview of Distributed Quantum Computing Emulators}
{\color{gray}\hrule}
\begin{center}
\textbf{Evaluation of Emulators for Distributed Quantum Circuit Execution}
\bigskip
\end{center}
{\color{gray}\hrule}

In this section, we review a wide selection of quantum emulators and simulators relevant to the execution of distributed quantum computing (DQC) protocols. Our choice of emulators is informed by the recent survey by Caleffi et al. \cite{caleffi2024distributed}, which provides a comprehensive overview of simulation tools, classifying them for distributed quantum computing into three main categories:
hardware-oriented (e.g., SQUANCH~\cite{squanchBartlett2018Aug}, NetSquid~\cite{netsquidCoopmans2021Jul}), protocol-oriented (e.g., SimulaQron~\cite{simulaqronDahlberg2018Sep}, SeQUeNCe~\cite{sequenceWu2021Sep}, QuiSP~\cite{quispSatoh2021Dec}, QuNetSim~\cite{qunetsimDiadamo2021Jun}), and application-oriented (e.g., NetQASM SDK~\cite{netqasmDahlberg2022Jun}, QNE-ADK~\cite{qneadk2025Aug}, DQC Executor~\cite{dqcexecutor2025Aug}). 

The emulators are evaluated using criteria relevant to DQC benchmarking tasks. These include not only native support for essential operations like teleportation or circuit-level logic enabling algorithms such as the Quantum Fourier Transform (QFT), but also deeper characteristics such as discrete-event simulation, realistic quantum networking, noise simulation.

In order to evaluate and compare different distributed quantum computing (DQC) emulators, it is essential to identify and classify the core features that enable or limit their suitability for benchmarking tasks. The emulators in this study are analyzed according to a set of capabilities that reflect their programmability, physical realism, scalability, and usability. This section reviews a set of emulators (and simulators) evaluated for their suitability in modeling DQC tasks. We focus  on their ability to simulate teleportation-based logic, node synchronization, noise models, and circuit‑level emulation. 

We group into three categories based on their importance to our benchmarking goals: required (\textbf{MUST}) features, important features that significantly enhance the emulator’s realism or flexibility, and useful features that improve practicality and performance. A summary of these capabilities is provided in Table \ref{tab:capabilities_summary}, and a detailed explanation of each feature follows below.

\begin{table}[htbp]
    \centering
    \label{tab:capabilities_summary}
    \renewcommand{\arraystretch}{1.2}
    \begin{tabular}{ll}
        \toprule
        \textbf{Category} & \textbf{Feature} \\
        \midrule
        \multirow{2}{*}{\textbf{MUST}} & Teleportation Support \\
         & QFT Gate Support \\
        \midrule
        \multirow{6}{*}{\textbf{IMPORTANT}} & Discrete-Event Simulation (DES) \\
         & Multi-Node Simulation \\
         & Quantum Network Modelling \\
         & Noise Modelling \\
         & Emulation Type \\
         & Parallel Execution \\
        \midrule
        \multirow{3}{*}{\textbf{USEFUL}} & Native Performance Metrics \\
         & Language Compatibility and Interface \\
         & Active Development and Documentation \\
        \bottomrule
    \end{tabular}
    \caption{Summary of Emulator Capabilities for DQC Benchmarking}
\end{table}

\textbf{MUST-have capabilities}. These features are considered essential for the emulators to be included in the benchmarking study:
\begin{itemize}
    \item Teleportation support (either qubit or gate level): The emulator must support qubit or gate teleportation between nodes. This can be either as a native operation or constructible from lower-level primitives such as Bell pair creation, mid-circuit measurements, feed-forward classical communication and conditional operations.
    \item QFT gate support: The emulator must provide native or constructible versions of the gates used in the Quantum Fourier Transform. These include Hadamard gates, parameterized Phase gates, and controlled Phase gates. For the teleportation step of the process, more standard gates such as CNOT, X and Z gates may also required.
\end{itemize}

\textbf{IMPORTANT capabilities}. These features are not strictly required, but are highly valuable for simulating realistic distributed quantum systems and flexibility:
\begin{itemize}
    \item Discrete-event simulation (DES):  allows accurate modeling of time-dependent behaviors, including synchronization between nodes, communication latency and delays. It is also essential to accurately model decoherence and other time-based noise effects. For an accurate emulator, this can be considered a MUST-have, but some non-DES emulators are included in the benchmarking to compare with others.
    \item Multi-node simulation: The ability to simulate more than two nodes is critical for assessing scalability and architectural behavior. Two-node simulation can be enough to compare basic scalability in the number of qubits and teleportations in this benchmarking test, but it is not enough for an actual DQC emulator.  
    \item Quantum network modelling: The emulator should support simulation of realistic quantum communication networks, including latency and message passing, photon losses, network topology and routing. 
    \item Noise modelling: Realistic quantum systems are noisy, and noise can completely break the performance of an algorithm. Emulators should allow noise in communication (e.g., depolarizing channels, photon loss) and noise in computation (e.g., gate errors, measurement noise). This could come from a fully open quantum system, a density matrix simulation, or some other technique to simulate noise.  
    \item Emulation type: The underlying simulation model affects expressiveness and performance. Common paradigms include statevector, density matrix, stabilizer or tensor network simulation.
    \item Parallel execution: The ability to run simulations across multiple threads, cores, or even machines increases scalability and reduces runtime.
\end{itemize}

\textbf{USEFUL capabilities}. These features are not critical for the simulation itself, but they are convenient to perform the benchmarking.
\begin{itemize}
    \item Native performance metrics: Built-in measurement of allocated memory and execution time. This is especially useful for automated benchmarking and analysis, but external tools can be used instead.
    \item Language compatibility and interface: Programming language support and API design affect usability. For instance, Python bindings or integration with quantum SDKs (like Qiskit or NetQASM SDK) may be essential for certain workflows.
    \item Active development and documentation: tools that are well-maintained with documentation and examples are more accessible for experimental comparison and fair benchmarking. 
\end{itemize}

Descriptions below highlight each emulator’s design goals, features, and limitations, and indicate whether they were included in our benchmarking effort. Only a few of them are included in the benchmarking test in the next section, as most of them lacked the necessary support for the chosen test. 

From the surveyed tools, we selected Qiskit Aer, SquidASM, Interlin-q, and SQUANCH for benchmarking, as they met the minimum requirements for simulating a gate-teleportation-based distributed inverse QFT, or provided an instructive comparison point despite known limitations. Other tools are briefly reviewed in Section 2.2, but were excluded from benchmarking due to missing capabilities or incompatible architectures.

\subsection{Included in Benchmarking}
\subsubsection{Qiskit Aer}

Qiskit Aer~\cite{aer2025Aug} is a high-performance circuit-level simulator provided by IBM as part of the Qiskit SDK~\cite{qiskit2025Aug}. Originally intended for monolithic circuit simulation, Qiskit Aer offers a flexible backend interface that includes statevector, density matrix, unitary, and stabilizer simulations. It is fully integrated with Qiskit’s transpiler stack and supports realistic noise models, making it suitable for algorithm development and validation under both ideal and noisy conditions.

While Qiskit Aer is not inherently a distributed quantum computing simulator, it can be adapted for DQC prototyping by manually decomposing circuits into separate node registers. Teleportation is not a native primitive but it can be built using a combination of standard operations, such as Bell pair generation, mid-circuit measurements, classical feed-fordward, and post-processing of measurement outcomes. This makes it suitable for validating logical correctness and for performance baselines, even if it does not account for inter-node latency or timing.

\textbf{Key features:}
\begin{itemize}
  \item Supports multiple backends: statevector, stabilizer, density matrix, unitary simulator.
  \item Integrates with Qiskit's transpiler for optimized gate decomposition and device-aware routing.
  \item Allows noise models including depolarizing, amplitude damping, readout error, and custom channels.
  \item Efficient execution for up to $\sim$30 qubits depending on hardware resources.
  \item Active development.
\end{itemize}

\textbf{Limitations:}
\begin{itemize}
  \item No notion of distributed nodes or classical message passing beyond manual simulation.
  \item No discrete-event timing or control over scheduling, cannot model synchronization or qubit lifetimes precisely.
  \item All teleportation and inter-node logic must be hand-coded using circuit primitives.
\end{itemize}

\textbf{Evaluation Context:} Used as a baseline for runtime and fidelity comparisons. It provides a best-case scenario against which true DQC emulators can be compared in terms of overhead, noise tolerance, and expressiveness.

\vspace{1em}
\subsubsection{SquidASM}

SquidASM~\cite{squidasm2025Aug} is a distributed quantum computing (DQC) simulation framework developed by QuTech (TU Delft), built directly on top of NetSquid~\cite{netsquidCoopmans2021Jul}, a modular discrete-event simulator for quantum networks. Unlike abstract protocol-level tools, SquidASM enables explicit modeling of both quantum and classical processes across networked quantum nodes with time-aware event scheduling. It was designed with realism and low-level control in mind, making it well-suited for algorithmic benchmarking and hardware-aware protocol design.

SquidASM introduces agents that act as quantum processing nodes. Each agent can perform quantum operations, schedule gates, initiate or consume entangled links, and exchange classical messages with other nodes. Classical and quantum control flows are explicitly coded by the user, allowing full freedom to coordinate teleportation protocols, circuit segmentation, and inter-node logic.

Unlike some emulators that abstract node behaviors, SquidASM requires users to explicitly handle gate timing, memory coherence lifetimes, communication latencies, and scheduling queues. This makes it particularly accurate for modeling protocol performance under real-world hardware constraints.

\textbf{Key Features:}
\begin{itemize}
  \item Full control over discrete-event scheduling and simulation time.
  \item Native support for teleportation, entanglement generation, swapping, and measurement-driven operations.
  \item Qubits are explicitly modeled with lifetimes, decoherence, and quantum memory constraints.
  \item Communication links simulate delays, noise, and entanglement fidelity degradation.
  \item Multi-node experiments can be executed with custom protocols and arbitrary topologies.
  \item Active development.
\end{itemize}

\textbf{Limitations:}
\begin{itemize}
  \item Requires manual definition of all quantum and classical protocols, including scheduling logic.
  \item Steep learning curve and limited documentation for advanced usage.
  \item Not fully open source, access requires a QuTech account and license
  \item Primarily oriented toward quantum internet protocols but adaptable for algorithmic DQC.
\end{itemize}

\textbf{Evaluation Context:} In our benchmarking study, SquidASM was the only platform able to support the complete distributed inverse QFT protocol with full phase encoding, teleportation, and classical feedback plus realistic discrete-event simulation.

\vspace{1em}
\subsubsection{Interlin-q}

Interlin-q~\cite{interlinqParekh2021Jun, interlinq2025Aug} is an application-oriented emulator developed by University of Naples Federico II and collaborators in the Quantum Internet Alliance (QIA). Intelin-q takes a monolithic circuit and partitions it for distributed execution using teleportation. It automates circuit partitioning and distributed execution over a QuNetSim~\cite{qunetsimDiadamo2021Jun} backend, which is used for simulating the quantum network.

It supports all the quantum gates necessary for the benchmarking test, with their documentation detailing a distributed version of the Quantum Phase Estimation as an use case, so the programability is straightforward. However, teleportation is handled natively and there was way to use the cat-entangler/disentangler optimizations used in Section~\ref{section3}, potentially increasing teleportation count and overhead as compared to other emulators.

\textbf{Key features:} 
\begin{itemize}
  \item Automates circuit partitioning and 
  \item Provides built-in support for teleportation and distributed execution using QuNetSim.
\end{itemize}
\textbf{Limitations:} prototype stage
\begin{itemize}
  \item Automatic partition is hardcoded to use basic gate-teleportation.
  \item No native noise modelling, should be implemented by hand with probabilistic gates.
  \item No active development.
\end{itemize}

\textbf{Evaluation Context:} Interlin-q was evaluated as a prototype DQC 
simulator that handles automatic circuit partitioning and teleportation. 
Its design is oriented toward high-level distributed execution using the 
QuNetSim backend. We included Interlin-q in our study primarily 
for qualitative comparison, noting its current status as an early-stage research 
tool rather than a mature emulator.

\vspace{1em}
\subsubsection{SQUANCH}

SQUANCH (Simulator for QUAntum Networks and CHannels)~\cite{squanchBartlett2018Aug} is a hardware-oriented emulator developed by the AT\&T Research Labs. It is a Python-based framework designed to simulate quantum networks through an agent-based model. It provides support for density matrix simulation of qubit systems and represents a middle ground between educational tools and low-level network modeling.

It uses classes such as QSystem, QStream, Agents and Channels. QSystem represents the quantum state of a multi-particle entangled system as a complex-valued density matrix. QStream represents a collection of disjoint (mutually unentangled) quantum systems, such as an ensemble EPR pairs. Finally, each agent in SQUANCH can send, receive, and process both quantum and classical information through user-defined channels. These channels simulate time delays and can introduce configurable noise. Internally, SQUANCH uses Python’s multiprocessing module and shared memory structures to simulate entangled states across nodes.

Although it provides a usable abstraction for distributed protocols, SQUANCH lacks timing granularity and does not implement discrete-event simulation. Moreover, it only supports density matrix representation, which limits its scalability in terms of memory footprint.

\textbf{Key features:}
\begin{itemize}
  \item Agent-based model with classical and quantum memory spaces, running in separate processes.
  \item Uses shared memory to manage entangled states between agents.
  \item Provides basic gate library, including teleportation primitives.
\end{itemize}

\textbf{Limitations:}
\begin{itemize}
  \item No discrete-event engine, timing is simulated via blocking/sleep delays.
  \item No native noise modelling, should be implemented by hand with probabilistic gates.
  \item No active development.
\end{itemize}

\textbf{Evaluation Context:} SQUANCH was tested for compatibility with DQC-style protocols. Due to its flexible API, it could support basic teleportation logic. We included SQUANCH for qualitative comparison.

\subsection{Other Tools Considered}

Several other emulators were reviewed but excluded from benchmarking due to missing features, incompatible architectures, or lack of gate-level programmability for the distributed inverse QFT task.  

\paragraph{NetSquid}  
A discrete-event simulator for quantum networks with nanosecond resolution and detailed modelling of noise, latency, and hardware effectsm developed by QuTech, and is part of the European Quantum Internet Alliance (QIA) software stack. NetSquid~\cite{netsquidCoopmans2021Jul} supports entanglement distribution, teleportation, and complex network topologies. However, it lacks a direct circuit-level interface and requires integration with higher-level tools (e.g., SquidASM, NetQASM) to execute gate-based algorithms. For this study, it was used indirectly via SquidASM but was not benchmarked on its own.  

\paragraph{SimulaQron}  
SimulaQron~\cite{simulaqronDahlberg2018Sep} is a virtual quantum network simulator, also developed by QuTech within the QIA to enable the application-level prototyping of distributed quantum protocols. It provides a network of virtual nodes, each running as a classical process, with simulated quantum registers that can be entangled across nodes. Quantum operations are delegated to pluggable local backends such as ProjectQ or QuTiP, while classical communication between nodes is handled through TCP sockets.

It does not implement discrete-event timing or detailed noise models, and circuit execution is limited so running the full distributed QFT is challenging. Furthermore, we expect it to scale poorly because of the dependence on the backends ProjectQ and QuTiP for the simulation. Furthermore, is is no longer actively developed, and has been deprecated by tools like NetQASM and SquidASM in the QIA ecosystem. While it was useful for early-stage protocol development, it is not appropriate for teleportation-intensive algorithms like our distributed QFT.  

\paragraph{NetQASM SDK}  
A high-level SDK and intermediate representation for distributed quantum applications. Another tool from QuTech and the QIA, NetQASM SDK abstracts hardware details and compiles quantum-classical programs for backends such as SquidASM or SimulaQron, which both use the low-level instruction set architecture NetQASM~\cite{netqasmDahlberg2022Jun}. Its value lies in portability and code reuse rather than direct simulation, but its performance and fidelity characteristics depend entirely on the backend used. For this reason, we focused on benchmarking SquidASM directly. 

\paragraph{SeQUeNCe}  
A modular discrete-event simulator developed at the University of Chicago and the Argonne National Laboratory (U.S.) for evaluating the performance of quantum communication networks, especially repeater chains and entanglement routing strategies~\cite{sequenceWu2021Sep}. It models link-layer behaviours, memory decoherence, and signal loss with precision, but its quantum logic support is limited to communication tasks. Lacking native teleportation gates or arbitrary quantum circuit execution, it was unsuitable for our distributed QFT benchmark.  

\paragraph{QuNetSim}  
An educational and prototyping tool for quantum communication protocols~\cite{qunetsimDiadamo2021Jun}. It provides an easy-to-use API for sending qubits, generating entanglement, and exchanging classical messages, but omits timing models, gate execution, and realistic noise. The absence of circuit-level control and teleportation-based computation support made it incompatible with our benchmarking goals, however it is used as a basis for Interlin-q for handling network control.  

\paragraph{DQC Executor}  
A framework for deploying distributed circuits using an OpenQASM input and YAML-based topology descriptions, mapping them to NetSquid simulations~\cite{dqcexecutor2025Aug}. While the tool is straightforward and has a low-entry barrier, it is in early stages and it is not actively developed. Furthermore, as it depends entirely on NetSquid it is more of an intermediate automation layer.

\paragraph{QNE-ADK}  . 
The Quantum Network Explorer Application Development Kit (QNE-
ADK)~\cite{qneadk2025Aug} is a high-level, application-oriented tool for setting up experiments and workflows in SquidASM and NetSquid. It automates scaffolding though command line commands, and abstracts away many simulator details. However, it is not a simulator by itself, depending completely on SquidASM and NetSquid. As previously mentioned, we focused on benchmarking SquidASM direcly. 

\paragraph{QuISP}  
A C++-based simulator designed for large-scale quantum communication and network routing analysis~\cite{quispSatoh2021Dec}, developed by the National Institute of Information and Communications Technology (NICT, Japan) in collaboration with Keio University. While it can handle massive numbers of entangled links with simplified error models, it lacks low-level gate fidelity tracking and teleportation logic inside nodes, preventing it from executing distributed quantum algorithms like QFT.  

\subsection{Emerging tools} 
We note that several new emulators and software tool for distributed or networked quantum computing have appeared recently, such as Qoala~\cite{van2025qoala} (another tool in the QIA family), NetMPI~\cite{cardama2025netqmpi}, or SimDisQ~\cite{zhang2025end}. These projects are promising and reflect the rapid evolution of the DQC software ecosystem, but they were identified only in the final stages of preparing this report and could not be evaluated within our benchmarking framework. We therefore leave their testing to future studies.

\subsection{Summary of Emulator Capabilities}

This subsection provides a consolidated overview of the capabilities of the 
distributed quantum computing emulators considered in this work. 
While these tools span different design goals, ranging from monolithic circuit 
simulation to full discrete-event modeling of quantum networks, they can be 
compared along a common set of criteria: teleportation support, multi-node 
execution, noise and network modeling, event-driven timing, and active 
development status. Table~\ref{tab:emulator_capabilities} summarizes these 
features and highlights which emulators were included in our benchmarking.

\begin{table}[ht]
\centering
\caption{Comparison of distributed quantum emulators by supported features. 
Symbols: \cmark = supported, $\sim$ = partially supported / requires user implementation / depends on other tools, \xmark = not supported.
The ``Included'' column indicates whether the emulator was benchmarked in this study.
The ``Active'' column reports the most recent year of public updates (documentation, release, or commit).}
\begin{tabular}{lccccccc}
\hline
\textbf{Emulator} & \textbf{Teleport.} & \textbf{QFT} & \textbf{Event Sim.} & \textbf{Net Model} & \textbf{Noise} & \textbf{Included} & \textbf{Active} \\
\hline
SquidASM     & \cmark & \cmark & \cmark & \cmark & \cmark & \cmark & 2025 \\
Qiskit Aer   & $\sim$ & \cmark & \xmark & \xmark & \cmark & \cmark & 2025 \\
Interlin-q   & $\sim$ & \cmark & \xmark & $\sim$ & \xmark & \cmark & 2021 \\
SQUANCH      & \cmark & \cmark & \xmark & $\sim$ & $\sim$ & \cmark & 2018 \\
NetSquid     & \cmark & $\sim$ & \cmark & \cmark & \cmark & \xmark & 2021 \\
SimulaQron   & \cmark & $\sim$ & \xmark & \cmark & $\sim$ & \xmark & 2021 \\
NetQASM SDK  & $\sim$ & $\sim$ & $\sim$ & $\sim$ & $\sim$ & \xmark & 2025 \\
SeQUeNCe     & \cmark & \xmark & \cmark & \cmark & \cmark & \xmark & 2025 \\
QuNetSim     & \cmark & \xmark & \xmark & \xmark & \xmark & \xmark & 2023 \\
DQC Executor & $\sim$ & $\sim$ & $\sim$ & $\sim$ & $\sim$ & \xmark & 2021 \\
QNE-ADK      & $\sim$ & $\sim$ & $\sim$ & $\sim$ & $\sim$ & \xmark & 2024 \\
QuISP        & \cmark & \xmark & \cmark & \cmark & $\sim$ & \xmark & 2025 \\
\hline
\end{tabular}
\label{tab:emulator_capabilities}
\end{table}

\newpage
\section{Benchmarking Methodology}
\label{section3}
{\color{gray}\hrule}
\bigskip

\subsection{Distributed Quantum Fourier Transform}

To benchmark the capabilities of distributed quantum emulators, we selected the Quantum Fourier Transform (QFT) algorithm as a testbed, implemented in a distributed fashion via gate teleportation. The QFT is a canonical quantum algorithm and a key subroutine in applications such as Shor’s factoring algorithm~\cite{shor1994} and quantum phase estimation~\cite{Kitaev_1997_QPE}. It performs a transformation analogous to the classical discrete Fourier transform, but with exponential speedup for certain structured inputs.

\begin{figure}[ht]
\centering
    \includegraphics[width=0.8\textwidth]{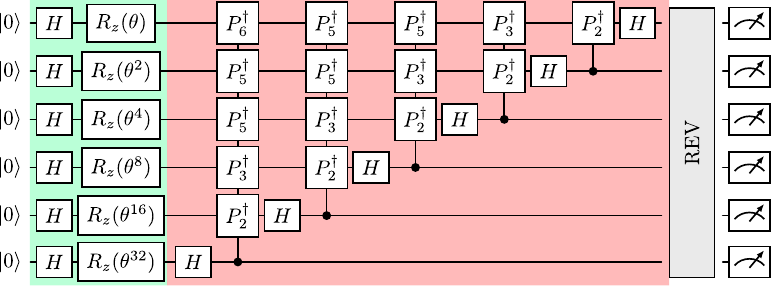}
\caption{Inverse QFT on 6 qubits (light red) with initial Fourier state preparation (light green). The state preparation consists of a Hadamard state followed by a sequence of phase gates. The QFT circuit applies Hadamard and controlled phase gates, where $P_k^{\dagger} = \mathrm{P}(e^{-2\pi i / 2^k})$. The reverse operator $REV$ (which stands for the standard swap gates at the end of the inverse QFT) can be realized by classical postprocessing of the measured bit strings, reducing considerably the entanglement cost~\cite{Chen2023qft}. }
\label{fig:qft-6q-fourier}
\end{figure}

The QFT transforms a computational basis state $\ket{x}$ into a superposition with relative phases:
\begin{equation}
    QFT\ket{x} = \frac{1}{\sqrt{N}} \sum_{k=0}^{N-1} e^{2\pi i xk/N} \ket{k},
\end{equation}
where $N = 2^n$ for an $n$-qubit system. It is typically implemented using Hadamard gates and controlled phase rotations, as shown in Figure~\ref{fig:qft-6q-fourier} resulting in high entanglement and global qubit connectivity, making it a stringent test of emulator capabilities. The inverse QFT is particularly well-suited for benchmarking because its output can be deterministically verified when applied to an input state prepared in the Fourier basis. Specifically, if the input encodes a phase $\theta$  via local $R_z$  rotations, the inverse QFT will yield a bitstring approximation of $\theta$ with high probability in the noiseless case.

\begin{figure}[ht]
\centering
    \includegraphics[width=0.8\textwidth]{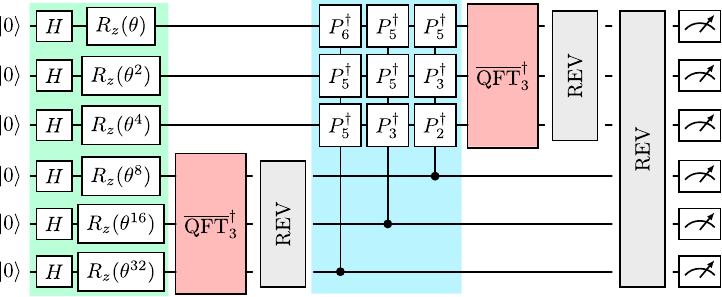}
\caption{By reordering commuting operations from Figure~\ref{fig:qft-6q-fourier} the inverse $\mathrm{QFT}$ can be rewritten into smaller, local inverse $\mathrm{QFT}$s (light red) and a block of controlled phase gradients (light blue). For a bipartite $\mathrm{QFT}^{\dagger}$ of $n$ qubits, we need $\frac{n}{2}$ EPRs to teleport the phase gradients. The $\overline{\mathrm{QFT}}^{\dagger}$ represents the usual inverse $\mathrm{QFT}$, without the reverse operator. The local $\mathrm{REV}$ operators can be moved forward by reversing the controls of the phase gradients, so that they can be applied classically in post-processing. }
\label{fig:qft-6q-fourier-2nodes}
\end{figure}

\begin{figure}[ht]
\centering
    \includegraphics[width=0.9\textwidth]{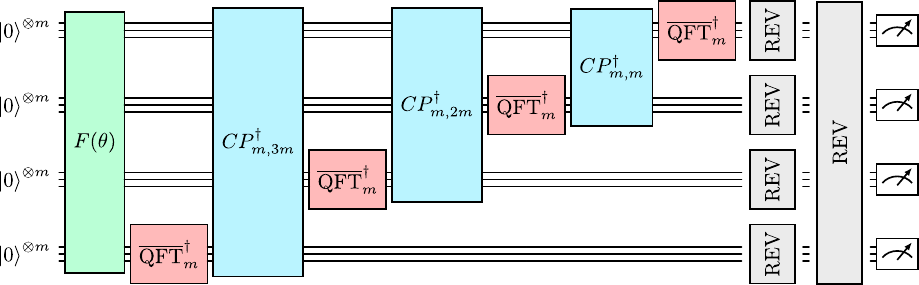}
\caption{Generalizing the circuit in Fig.~\ref{fig:qft-6q-fourier-2nodes} to an arbitrary number of nodes $k$, each one with $m$ qubits, so the whole computation involves $n = km$ qubits. Here $F(\theta)$ is the (local) preparation of the Fourier state, ${\mathrm{QFT}}^{\dagger}_{m}$ is the (local) $n$-qubit inverse QFT, $CP_{a,b}^{\dagger}$ is the block of controlled phase gradients with $a$ control qubits and $b$ target qubits. Because we pushed the $\mathrm{REV}$ blocks to the end of the circuit as classical post-processing, we need to reverse the order of controls in the controlled phase gradient block. Each phase gradient block can be executed remotely through the use of telegate. The total number of needed EPRs becomes $m\binom{k}{2}=\frac{mk(k-1)}{2}$. If all of the individual $CP$ rotations were to be executed remotely, this number would grow to $\binom{mk}{2}$, that is, quadratic not only in the number of nodes but also the number of qubits per node. Assuming full parallelization of quantum operations, the total number of computing blocks is $2k-1$. Each block of phase gradients involves $m$ layers of phase gates and each ${\mathrm{QFT}}^{\dagger}_{m}$ involves $2m-1$ layers of gates. While the ${\mathrm{QFT}}^{\dagger}_{m}$ are occurring in one node, the rest are executing phase gradients on the rest, with minimal idle time. Measurements can be performed as soon as qubits have finished their gates. }
\label{fig:qft-4nodes}
\end{figure}

To distribute the QFT across quantum nodes, we employ a recursive factorization of the circuit. Each node handles a local inverse QFT ($\mathrm{QFT}^\dagger_m$) on its $m$ local qubits, while inter-node phase gradients are handled via controlled operations teleported using entangled Bell pairs (EPRs), as shown in Figure~\ref{fig:qft-6q-fourier-2nodes} for only two nodes. For more than two nodes, we follow a recursive structure shown schematically in Figure~\ref{fig:qft-4nodes}, where each controlled gradient is sent across nodes using a single EPR per control qubit and receiving node via \textit{cat-entangler/disentangler} circuits~\cite{Yimsiriwattana2005,neumann2020imperfect}.

An initial Fourier basis state is prepared locally by each node using Hadamard gates and parameterized $R_z$ rotations of the form $\theta 2^{j}$. The inverse QFT then recovers the encoded phase, with measurements producing an $n$-bit approximation of $\theta$.

The inter-node operations are implemented through gate teleportation. Specifically, we use the \textbf{telegate} model, in which a control qubit is teleported to the target node using an EPR pair and entangled with a cat-entangler. This allows a series of local controlled phase gates to be applied using the teleported control. A final cat-disentangler circuit restores unitarity and coherence~\cite{Yimsiriwattana2005}.

This method enables the simultaneous teleportation of all controlled-phase gates sharing the same control qubit within a QFT layer. In a fully connected QFT circuit over $N = m k$ qubits (with $k$ nodes of $m$ qubits each), every qubit ideally interacts with all others, leading to $\binom{mk}{2}$ two-qubit gates in total. Each inter-node controlled gate would require a separate teleportation, resulting in a number of required EPR pairs that scales as $\mathcal{O}(m^2 k^2)$. 

By grouping all gates originating from the same control qubit through the \emph{cat-entangle} and \emph{cat-disentangle} procedures, only one teleportation is needed per control qubit and remote node. Consequently, the total number of required EPR pairs is reduced from $\binom{mk}{2}$ to approximately $m \binom{k}{2}$, achieving an overall scaling of $\mathcal{O}(m k)$. This linear dependence on both the number of nodes and the number of qubits per node represents a significant improvement over the naïve quadratic scaling in $m$ and $k$.

\subsection{Teleportation-Free Alternatives}

We note that an alternative to quantum teleportation is to use deferred measurement and classical control to implement dynamic circuits~\cite{Baumer2024dynamic}. In this version, measurements are performed early and classical outcomes are transmitted to other nodes to control local gates. This avoids quantum communication entirely and resembles the circuit structure of iterative quantum phase estimation~\cite{Whitfield2011iqpe}. Although we focus on gate-teleportation-based QFT in this benchmark, dynamic circuit approaches may offer useful comparisons for future work.

\subsection{Performance Metrics}

To assess emulator performance, we measure:

\begin{itemize}
    \item \textbf{Execution time:} total time to complete the simulation or emulation of the QFT circuit.
    \item \textbf{Memory usage:} peak memory required during execution, including any inter-node state tracking.
    \item \textbf{Fidelity:} similarity of the final state to the ideal QFT output measured via bitstring distributions (classical fidelity $F=\sum_i \sqrt{p_i q_i}$). Used as a sanity check to make sure the emulator is working as intended.
\end{itemize}

All metrics are evaluated as a function of the number of qubits $n$ and the number of nodes $k$. For a strictly statevector emulator, both memory and execution time are expected to grow exponentially with $n$, as the size of the Hilbert space (and thus the statevectors and unitary matrices) grows accordingly. The scaling with the number of nodes $k$ may uncover the actual capabilities of the emulator for distributed quantum computing. The number of Bell pairs (created and consumed), intermediate measurements and classical messages is expected to grow linearly with $k$ for the test presented in this manuscript. 

\subsection{Challenges in Benchmarking}

Benchmarking emulators poses several practical and theoretical challenges:

\begin{itemize}
    \item \textbf{Heterogeneous APIs:} Each emulator uses a different programming interface, circuit model, and node coordination method.
    \item \textbf{Heterogeneous Environments:} Each emulator has different dependencies, so they need to be installed in isolated environments (e.g. conda) and loaded separately for each job.
    \item \textbf{Gate and message handling:} Emulators vary in their handling of message passing, network latency, mid-circuit measurements, and classical control.
    \item \textbf{Resource tracking:} Not all emulators expose runtime or memory statistics, so custom monitoring has been used for this. We use basic Python instructions from the `time' and `tracemalloc' python libraries.  
    \item \textbf{Circuit fidelity:} Emulators may implement approximate or symbolic representations of nonlocal gates.
    \item \textbf{Scalability:} Some emulators fail or degrade significantly for moderate node counts or qubit depths due to serialization or backend constraints. 
\end{itemize}

To address these, we structured our benchmarking suite to abstract the benchmark task and use emulator-specific adapters for execution.

\subsection{Implementation, Workflow and Automation}

All simulations were executed on the QMIO HPC cluster at CESGA~\cite{Cacheiro2025May}, a hybrid classical-quantum computing system. Each compute node of QMIO consists of 2x Intel Xeon Ice Lake 8352Y (\textbf{ILK}) with 32 cores each (64 cores per node), with 256Gb of RAM and 1Tb of main memory per node. Jobs were submitted via the SLURM workload manager, ensuring consistent resource allocation and scheduling.

For each emulator, simulations were run in isolated conda environments to manage dependencies and avoid version conflicts. We used fixed software versions (listed below) to ensure reproducibility. The following software versions were used:
\begin{itemize}
    \item Qiskit Aer v0.17.1, with Qiskit v2.1.1 and Python v3.10.16
    \item SquidASM v0.13.4, with NetSquid v1.1.7 and Python v3.8.20
    \item Interlin-q v0.0.1.post5, with QuNetSim v0.1.3.post1 and Python v3.10.18
    \item SQUANCH v1.1.0, with Python v3.10.16
\end{itemize}

Additional package dependencies (e.g., \texttt{numpy}, \texttt{scipy}, \texttt{matplotlib}) were managed through 
\texttt{conda}, ensuring reproducibility. Full environment YAML files are provided in the supplementary GitHub repository.

To manage the large-scale parameter sweeps required for benchmarking, we developed an automated workflow. A central configuration file (\texttt{sweep\_params.yaml}) defines the ranges for qubit counts (n), node counts (k), phase angles ($\theta$), and target emulators. A Python script (\texttt{job\_creator.py}) parses this configuration and generates SLURM job submission scripts for each emulator and parameter combination by filling a template (\texttt{slurm\_template\_multinode.sh}). This ensures consistent resource allocation and environment setup across all runs. Jobs are submitted either collectively using \texttt{job\_submit\_all.sh} or individually via \texttt{job\_submit\_one.sh} for targeted testing. This modular approach facilitates reproducibility and allows for easy extension of the benchmarking suite.

First we assigned one single core per simulation, with 4Gb of RAM per core, and later parallelization behavior was assessed by varying the number of allocated \textbf{ILK} cores from 1 to 8. At no point of the simulations there was inter-node communication. Each job was allocated up to 24 hours of walltime, with memory limits scaled according to expected Hilbert space size.

Directory structure of the benchmarking suite. The central configuration (\texttt{sweep\_params.yaml}) drives automated job generation via \texttt{job\_creator.py}, which populates per-emulator jobs/ directories. Results are collected in results/ and parsed, analyzed and visualized using \texttt{plot\_results.py}
\dirtree{%
.1 dqc-emulator-benchmark/.
.2 configs/sweep\_params.yaml.
.2 scripts/*.py, *.sh.
.2 emulators/.
.3 qiskit\_aer/.
.4 jobs/, results/, scripts/*.py, env/*.yml.
.3 squidasm/.
.4 jobs/, results/, scripts/*.py, env/*.yml.
.3 interlinq/.
.4 jobs/, results/, scripts/*.py, env/*.yml.
.3 squanch/.
.4 jobs/, results/, scripts/*.py, env/*.yml.
.2 plots/ (auto-generated).
}


\newpage
{\color{gray}\hrule}
\begin{center}
\section{Scalability Tests and Results}
\textbf{Performance evaluation of distributed QFT implementations}
\bigskip
\end{center}
{\color{gray}\hrule}

\subsection{Description of the scalability test}

To evaluate the performance of emulators for distributed quantum computing (DQC), we executed the distributed inverse Quantum Fourier Transform (QFT) on Fourier basis states of increasing size. The benchmark targets execution time, memory consumption, and fidelity as a function of both the number of qubits and the number of computing nodes.

Distributed emulation introduces overheads due to teleportation of nonlocal gates, classical coordination, and synchronization. In the QFT, entangling gates between all qubit pairs make it a stress test for any distributed architecture. Our implementation uses gate teleportation via EPR pairs, and partitions the QFT across $k$ nodes of $m$ qubits each, with $n = km$. As shown in Figure~\ref{fig:qft-4nodes}, each node executes a local inverse QFT subroutine and participates in remote gate execution through cat-entangler and cat-disentangler primitives.

The theoretical resource requirements grow exponentially with the number of qubits: the Hilbert space dimension doubles per qubit, stressing both memory and compute time. Teleportation introduces further costs proportional to the number of inter-node entangling gates. We used the exact QFT algorithm (no gate truncation), which results in $\mathcal{O}(m k^2)$ teleported gate blocks.

An ideal distributed emulator would parallelize across nodes, pipeline computation and communication, and dynamically manage memory and EPRs to maintain high fidelity with efficient resource use. Our benchmarking tests these abilities directly.

\subsection{Methods}

The benchmark algorithm was implemented in three emulators: Qiskit Aer, SquidASM, and SQUANCH. Each implementation followed a common high-level structure: preparing a Fourier state with a phase $\theta$, executing the inverse QFT via teleportation across nodes, and measuring output qubit states. The results were compared to a monolithic (non-distributed) reference circuit using fidelity and resource metrics.

The test parameters are summarized in Table~\ref{tab:params}. We varied:
\begin{itemize}
    \item The total number of logical qubits $n$, from 4 to 20.
    \item The number of nodes $k$, from 1 to 8. The number of logical qubits $n$ does not need to be a multiple of $k$, except for Interlin-q.
    \item The Fourier phase angle $\theta = \{0, 1/3, 2/3\}$. We explore several non-integer angles and average the metrics for them.
    \item Execution noise models, when available. 
    \item Emulator-specific configurations (message logging, qubit partitioning, etc).
\end{itemize}

Fidelity between monolithic and distributed runs was computed from measurement outcomes using classical probability overlap. This does only make sense for noiseless emulation. 

\subsection{Results and Emulator Comparisons}

\paragraph{Qiskit Aer.}  
Qiskit Aer is not a native DQC emulator but supports gate teleportation through manual programming of classical control flow. We implemented teleportation circuits explicitly, resetting and reusing ancillary qubits for EPRs. Despite the overhead, Qiskit performed well up to 16–18 qubits and 4 nodes. Its use of optimized statevector simulation likely accounts for its speed in the noiseless case. However, memory and time scale poorly with increased nodes, and it lacks primitives for quantum network noise (e.g., entanglement fidelity or latency modeling). Furthermore, message passing is entirely implicit, limiting visibility into communication overhead.

\paragraph{SquidASM.}  
SquidASM is built for distributed execution, compiling circuits into NetSquid simulations. The network topology, qubit layout, and link definitions are programmable via YAML configuration files. Although native support for certain gates (e.g., controlled-phase) was missing, they were constructed from basic gates. SquidASM proved reliable and accurate, producing high-fidelity results even with increasing nodes. Execution times were longer than Qiskit for small problems but scaled more favorably with node count, suggesting better parallelization and concurrency. Importantly, SquidASM simulates full quantum state evolution with noise and messaging delays, making it well-suited for protocol evaluation.

\paragraph{Interlin-q.}
Interlin-q provides a higher-level abstraction for distributed quantum computing, automating the circuit partitioning and teleportation processes across multiple nodes. It builds on the QuNetSim communication framework and handles both circuit cutting and message scheduling internally. In our tests, Interlin-q correctly executed small-scale distributed inverse QFTs but exhibited faulty fidelities and poor scaling for larger systems, worsening with the number of nodes. Furthermore, the tool requires the same number of qubits per node, which constrained circuit partitioning and limited automation. Moreover, its automatic partitioner does not implement optimized distribution techniques such as the \emph{cat-entangle} and \emph{cat-disentangle} grouping, leading to redundant teleportations and communication overhead. Execution time and memory usage both grew exponentially with the number of qubits and nodes, and some large configurations failed to complete. While conceptually elegant, the implementation remains at a prototype stage and requires significant optimization for accurate and scalable emulation.

\paragraph{SQUANCH}
SQUANCH offers an intuitive distributed programming model but lacks a discrete-event engine. It failed to synchronize teleportation properly, causing fidelity to drop sharply as circuit depth or node count increased. While it showed promising scaling both in time and memory, specially with larger number of nodes, its scaling for monolithic emulation was very steep. While SQUANCH appears to parallelize efficiently its computation between nodes, it does struggle with large node sizes. Known issues with its coherence tracking and message handling were evident in our tests.  Furthermore, SQUANCH's simulation did not include any version of noise or realistic behavior. These findings match prior observations that SQUANCH struggles with multi-hop entanglement and gate synchronization.

\subsection{Resource Metrics and Trends}

\begin{figure}[htbp]
    \centering
    \includegraphics[width=1.0\textwidth]{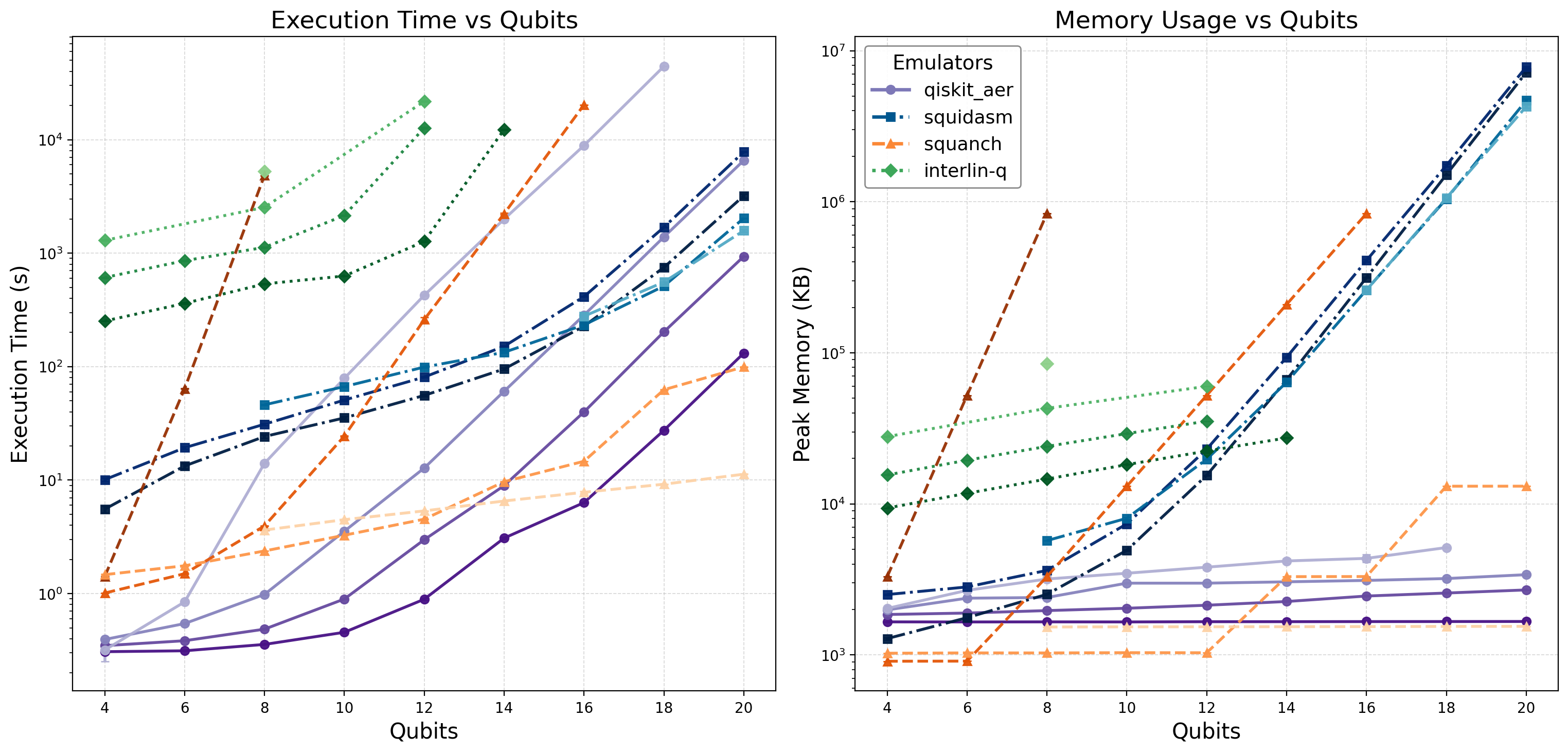}
    \caption{Execution time and peak memory usage as a function of the number of qubits, for different numbers of nodes and emulators. Each emulator is represented by a colour gradient: the darkest curve corresponds to single-node (monolithic) execution, while progressively lighter shades indicate increasing node counts.}
    \label{fig:scaling1}
\end{figure}

\begin{figure}[htbp]
    \centering
    \includegraphics[width=0.7\textwidth]{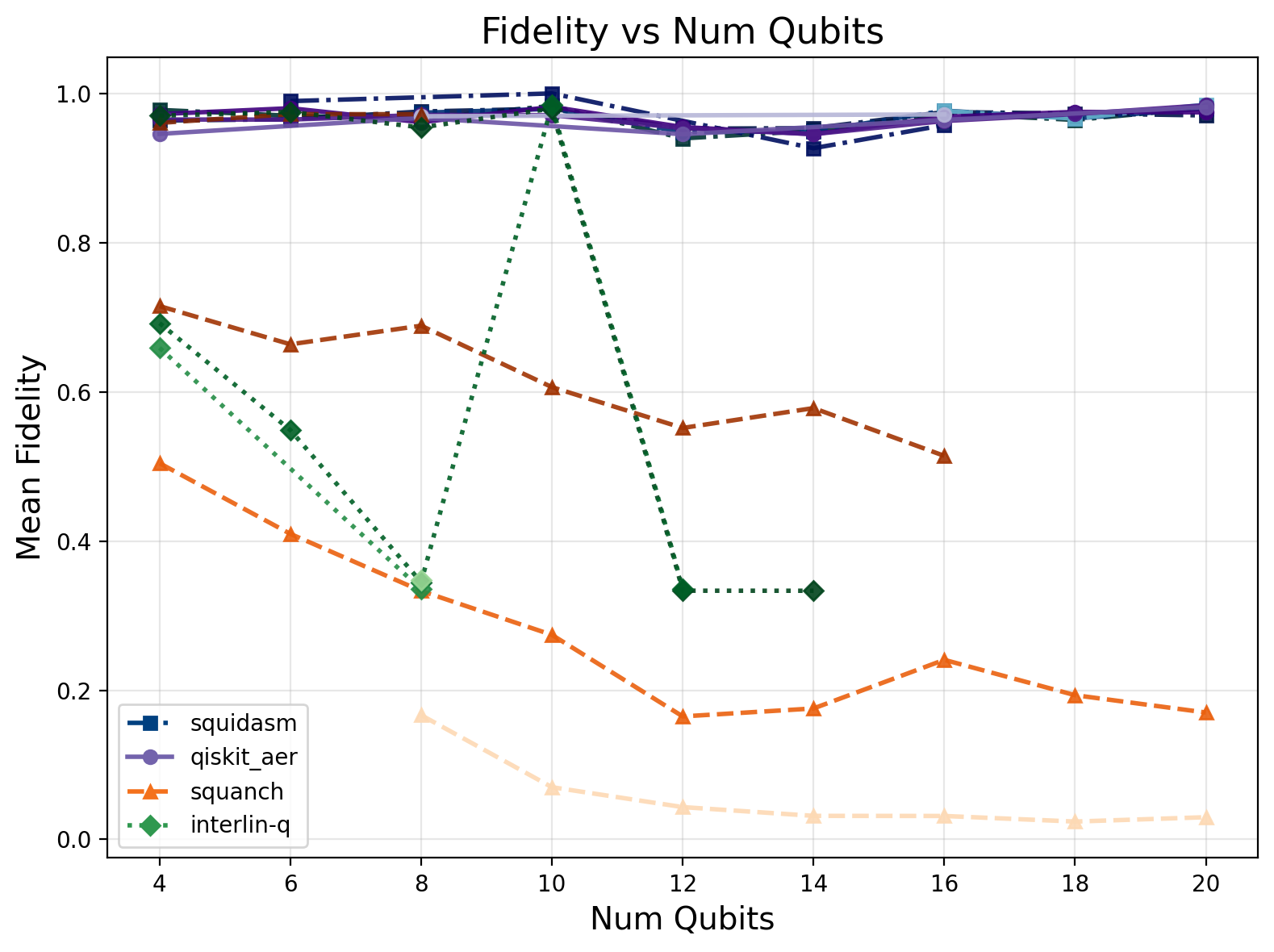}
    \caption{Classical fidelity of noiseless execution as a function of the number of qubits, for different numbers of nodes and emulators. As in \ref{fig:scaling1}, lighter shades of the same colour indicates increasing node counts. }
    \label{fig:scaling2}
\end{figure}

\begin{itemize}
    \item \textbf{Execution time} increases exponentially with qubit count, as seen in Fig.~\ref{fig:scaling1}, which is to be expected. Qiskit Aer is faster for small sizes, but SquidASM's performance improves relative to Qiskit as node count increases. Interlin-q is slower than both, increases with node number, and shows a growing slope for higher qubit numbers. For SQUANCH however, both metrics improve dramatically for the larger number of nodes. Most of the experiments with interlin-q and SQUANCH did not even finish in the allocated time.
    \item \textbf{Memory usage} similarly grows with circuit size, however as seen in the second panel of Fig.~\ref{fig:scaling1} Qiskit Aer memory usage grows sub-exponentially with qubit count, unlike SquidASM, SQUANCH and Interlin-q, which both follow exponential trends. SquidASM shows more stable memory behavior with the number of qubits, with a steady exponential increase. However, increasing the number of nodes still increases Qiskit Aer's peak memory consumption, while parallel scalability is better handled in SquidASM, which shows flat or even decreasing time with more nodes in some regimes, unlike Qiskit. Interlin-q shows an exponential growth in both qubit number and node number, while SQUANCH shows great efficiency in memory with more, smaller nodes.
    \item \textbf{Fidelity of noiseless execution} shown in Fig.~\ref{fig:scaling2} stays constant for both SquidASM and Qiskit Aer across all tested qubit counts and node numbers, indicating that their distributed execution models preserve the logical structure of the circuit with negligible numerical error (up to shot noise variance). In contrast, Interlin-q and SQUANCH exhibit substantially reduced fidelities that degrade with increasing problem size. For these emulators, fidelity decreases more rapidly as either the number of qubits or the number of nodes grows, even in the absence of noise. 
\end{itemize}

\begin{table}[h]
\centering
\caption{Summary of parameters used in the benchmarking experiments.}
\label{tab:params}
\begin{tabularx}{\textwidth}{|l|X|l|}
\hline
\textbf{Parameter} & \textbf{Description} & \textbf{Values Used} \\
\hline
\texttt{Num Qubits} & Number of qubits in the monolithic circuit & 4, 6, 8, 10, 12, 14, 16, 18, 20 \\
\hline
\texttt{Nodes} & Number of distributed nodes & 1, 2, 4, 8 \\
\hline
\texttt{Theta} & Phase parameter for Fourier state preparation & 0.0, 0.33, 0.66 \\
\hline
\texttt{Shots} & Number of circuit repetitions & 100 \\
\hline
\texttt{Backends} & Simulators used for execution & \texttt{squidasm}, \texttt{qiskit\_aer}, \texttt{interlin-q}, \texttt{squanch} \\
\hline
\texttt{Topology} & Distribution scheme among nodes & All-to-all connectivity \\
\hline
\texttt{Partition Strategy} & Distribution of qubits per node & Evenly partitioned, remainder in last node \\
\hline
\texttt{Output Metrics} & Measured performance & Execution time, memory usage, fidelity \\
\hline
\end{tabularx}
\end{table}

\newpage
{\color{gray}\hrule}
\begin{center}
\section{Discussion and Conclusions}
\bigskip
\end{center}
{\color{gray}\hrule}

\subsection{Discussion}

Our benchmarking of distributed quantum Fourier transform (QFT) implementations across multiple emulators highlights both the current capabilities and the limitations of existing distributed quantum computing (DQC) emulation platforms. The results provide valuable insights into the trade-offs between scalability, accuracy, and resource usage when simulating quantum algorithms in a distributed fashion.

Each emulator illustrates a different point in the design space of DQC simulation: Qiskit Aer as a flexible monolithic baseline, SquidASM as a purpose-built discrete-event simulator, Interlin-q as a partitioning prototype, and SQUANCH as a lightweight conceptual tool

Qiskit Aer shows strong performance as a monolithic simulator adapted for DQC through manual circuit partitioning. However, it has no native notion of a quantum network, as teleportation and classical coordination must be implemented by hand, and there is no built-in support for communication latency, EPR fidelity, or other network-level imperfections. Memory growth in our tests was unexpectedly modest, suggesting Aer employs a memory-efficient representation (e.g., decision diagrams) rather than a pure statevector or density matrix. Runtime benefits from Aer’s inherent parallelization, but execution time still grows rapidly with the number of nodes due to teleportation overhead.

SquidASM offers the most complete feature set for realistic DQC emulation, with full discrete-event simulation, network noise modeling, and accurate synchronization. Both time and memory usage scale exponentially with total qubits, as expected, but distributing the same number of qubits across more nodes can improve scaling, indicating efficient memory management and concurrent execution across agents. In contrast to Qiskit Aer, adding CPU cores did not yield substantial performance gains. While SquidASM is primarily oriented toward quantum internet protocols, it can be adapted for algorithmic DQC benchmarking. A major non-technical limitation is that it is not open source, requiring authentication through QuTech’s portal to obtain a license.

Interlin-q automates circuit partitioning and teleportation scheduling, but its scaling is strictly exponential and worsens with increased node count. The tool requires equal qubit counts per node, limiting flexibility in partitioning strategies and complicating automation. More critically, its partitioning does not implement our cat-entangler/cat-disentangler optimization, likely resulting in many more teleportations than necessary. This inefficiency, combined with its rigid architecture, constrains its suitability for large-scale DQC algorithm emulation.

SQUANCH provides an intuitive API for describing distributed quantum protocols, but its lack of a discrete-event simulation engine and limited synchronization support lead to severe fidelity degradation as qubit count or node count increases. In our tests, teleportation could not be consistently synchronized at larger scales, and execution times were significantly higher than with other platforms. While its architecture is suitable for conceptual exploration and education, these limitations make it unsuitable for quantitative benchmarking alongside the other emulators in this study. For this reason, we present SQUANCH’s results separately from the main performance figures. 

From a methodological perspective, we find that emulators vary significantly in how they handle:
\begin{itemize}
    \item Mid-circuit measurements and classical feedforward;
    \item Distributed entanglement and EPR pair management;
    \item Scheduling and time synchronization across nodes;
    \item Resource reporting (e.g., message statistics or memory snapshots).
\end{itemize}
These differences affect not only simulation performance but also the reproducibility and interpretability of results across platforms.

The recursive distributed QFT algorithm used in our benchmark proves to be a useful stress test, as it combines local computation with teleportation-intensive remote gate blocks. It exercises a variety of core features required in distributed systems, such as parameterized gate control, qubit routing, and classical coordination between nodes.

It is important to note that the distribution strategies implemented in different emulators 
are not always equivalent. In particular, tools such as Interlin-q perform automatic gate 
teleportation for every two-qubit gate whose operands reside on different nodes, which can 
introduce a significant communication overhead. In contrast, our custom implementation 
introduces an optimized procedure based on \emph{cat-entangle} and \emph{cat-disentangle} 
operations, which groups all controlled gates sharing the same control qubit into a single 
teleportation step. This reduces the number of required EPR pairs from $\binom{mk}{2}$ to 
$m \binom{k}{2}$, effectively improving the scaling of communication resources from 
$\mathcal{O}(m^2 k^2)$ to $\mathcal{O}(m k)$. 

While this optimization may not be directly comparable to the default behavior of other 
emulators, it represents a general and transferable approach for efficiently distributing 
QFT circuits in a multi-node environment. Consequently, we include this variant as an 
illustration of how algorithm-aware scheduling can significantly reduce the cost of 
distributed quantum simulations.

\subsection{Conclusions}

Our study demonstrates that distributed quantum computing emulators are becoming increasingly capable, but there remains substantial variability in feature support, scalability, and performance. The key takeaways are:

\begin{itemize}
    \item \textbf{Emulation of distributed quantum algorithms is possible}, but requires careful algorithm decomposition and tool-specific adaptations.
    \item \textbf{Resource efficiency varies widely across platforms}: low-level discrete-event emulators like SquidASM offer superior memory and runtime scalin, while higher-level tools (e.g.\ Interlin-q, SQUANCH) show reduced fidelity or fail at larger problem sizes.
    \item \textbf{Fidelity remains a concern at large scales}, especially in platforms that abstract communication or do not natively support gate teleportation.
    \item \textbf{Distributed execution is not yet standard} in many general-purpose simulators. Tools like Qiskit Aer can be extended for DQC, but their internal models are not optimized for teleportation and cross-node messaging.
\end{itemize}

More broadly, our results highlight a gap in the current ecosystem: to the best of our knowledge, no toll simultaneously emulates physically distributed quantum processors \emph{and} supports distributed-memory classical simulation. Such a \emph{distributed distributed} emulator would allow scalable studies of multi-node architectures and network-aware algorithms, but remains technically challenging and an important direction for future work.

As quantum hardware continues to evolve, emulation and simulation will remain crucial for developing distributed protocols and system architectures. Our benchmarking methodology and test circuits can serve as a foundation for future assessments of DQC platforms.

\subsection{Limitations and Future Work}

The main limitations of our study are the relatively small number of emulators tested and the fixed benchmark algorithm. Due to time and engineering constraints, only four emulators were fully evaluated, though many more were reviewed in the emulator overview section. NetQASM SDK, while promising as a DQC SDK, was not benchmarked due to compatibility issues between its SquidASM and Simulaqron backends. Simulaqron itself lacks support for general quantum circuits and was excluded. Other emulators (SeQUeNCe, QuNetSim, QuISP) are focused on network-level simulations and  were excluded due to lack of native gate-level programmability, circuit construction primitives, or missing teleportation support.

In addition, our analysis did not explore several system-level performance aspects. We did not perform a detailed study of multithreading, GPU acceleration, or backend-specific optimizations, even though these can strongly influence runtime and memory behavior. For example, Qiskit Aer supports GPU backends, but such capabilities were outside the scope of this work. Furthermore, we did not examine compiler optimizations, circuit re-writing strategies, or teleportation scheduling policies beyond our hand-crafted cat-entangle/disentangle approach. These factors can significantly alter both communication overhead and global resource usage.

Our results also reflect default simulation modes, which could impact performance. For example, Qiskit Aer can switch between statevector and density-matrix mode, or use a stabilizer 
simulator for Clifford-dominated circuits, while SquidASM can configure different physical- and link-level models within NetSquid. These choices can significantly affect runtime, memory usage, and even fidelity, but were beyond the scope of the present work. Exploring alternative formalisms for the emulators (e.g., stabilizer, density-matrix, or tensor-network engines) is an important direction for future benchmarking. 

Finally, our benchmark focused exclusively on an all-to-all algorithmic connectivity pattern, as induced by the inverse QFT. This topology does not exercise routing, path selection, or multi-hop entanglement distribution, and therefore does not reveal the strengths or weaknesses of emulators in more realistic network scenarios. For instance, event-driven simulators such as SquidASM or SeQUeNCe are explicitly designed to model routing and link-level contention. Exploring performance under constrained network topologies therefore remains an important direction for future evaluation.

In future work, we plan to extend this benchmarking framework to:
\begin{itemize}
    \item Include additional emulators as they mature and expand their APIs.
    \item Benchmark other distributed algorithms such as quantum teleportation networks, distributed Grover's search, or dynamic QKD schemes.
    \item Explore noise models, link-level imperfections, and error-injection for robustness testing
    \item Evaluate different simulation formalisms within each emulator.
    \item Log and analyze quantum/classical message traffic and latency more explicitly.
    \item Study the influence of classical performance features such as multithreading or GPU acceleration on the scalability of emulation.
    \item Evaluate emulators under non-trivial network topologies (line, grid, tree, and general graphs), including routing, entanglement swapping, multi-hop scheduling, and path-dependent latency.
\end{itemize}

Finally, we will release the full benchmarking suite, including scripts, configuration templates, and plotting routines, openly on GitHub, enabling the community to reproduce and expand upon our results.

\section*{Acknowledgments} 
{\small This work was supported by MICINN through the European Union NextGenerationEU recovery plan (PRTR-C17.I1), the Galician Regional Government through ``Planes Complementarios de I+D+I con las Comunidades Autónomas'' in Quantum Communication. The authors also acknowledge Galicia Supercomputing Center (CESGA) for providing access to the FinisTerrae III supercomputer with financing from the Programa Operativo Plurirregional de España 2014-2020 of ERDF, ICTS-2019-02-CESGA-3. This research project was made possible through the access granted by the Galicia Supercomputing Center (CESGA) to its Qmio quantum computing infrastructure with funding from the European Union, through the Programa Operativo Galicia 2014-2020 of ERDF--REACT EU, as part of the European Union’s response to the COVID-19 pandemic.
{\small During the preparation of this work the authors have used AI tools to revise grammar and readability. The authors take full responsibility for the content of the publication.}
\newpage
\bibliographystyle{bibstyle}
\bibliography{references}
\end{document}